# Topology control of human cells monolayer by liquid crystal elastomer

Taras Turiv[1,2]*, Jess Krieger[3], Greta Babakhanova[1,2], Hao Yu[1,2],
Sergij V. Shiyanovskii[2], Qi-Huo Wei[1,2,4], Min-Ho Kim[3,5], Oleg D. Lavrentovich[1,2,4]*

**Affiliations:**

[1]Chemical Physics Interdisciplinary Program, Kent State University, Kent, OH, 44242, USA

[2]Advanced Materials and Liquid Crystal Institute, Kent State University, Kent, OH, 44242, USA

[3]School of Biomedical Sciences, Kent State University, Kent, OH, 44242, USA

[4]Department of Physics, Kent State University, Kent, OH, 44242, USA

[5]Department of Biological Sciences, Kent State University, Kent, OH, 44242, USA

*Authors for correspondence: tturiv@kent.edu, olavrent@kent.edu

**Abstract:** Biological cells in living tissues form dynamic patterns with local orientational order and topological defects. Here we demonstrate an approach to produce cell monolayer with the predesigned orientational patterns using human dermal fibroblast cells (HDF) placed onto a photoaligned liquid crystal elastomer (LCE). The alignment of cells is caused by anisotropic swelling of the substrates in contact with the aqueous cell growth medium. The patterns predesigned in the LCE cause a strong spatial variation of cell phenotype (evidenced by shape variations), their surface density and number density fluctuations. The concentration of cells is significantly higher near the cores of positive-strength defects as compared to negative-strength defects. Unbinding of defect pairs intrinsic to active matter is suppressed by anisotropic surface anchoring. The geometry of arrays allows one to estimate the elastic and surface anchoring characteristics of the tissues. The demonstrated patterned LCE approach could be used to control the collective behavior of cells in living tissues, cell differentiation, and tissue morphogenesis.

**Introduction**

Biological tissues formed by cells in close contact with each other often exhibit orientational order, caused by mutual alignment of anisometric cells (*1-5*). The direction of average orientation, the so-called director $\hat{\mathbf{n}}$, varies in space and time, producing topological defects called disclinations (*6, 7*). These defects move within the tissue and play an important role in compressive-dilative stresses and effects such as extraction of dead cells (*6*). Finding means to design a tissue scaffold for living cells with orientational order and to control the topological type and dynamics of defects is of a significant importance in the biomedical field and for our understanding of how living matter can be manipulated. A number of research groups demonstrated a spectacular progress in production of ordered assemblies of cells at lithographically fabricated substrates (*8, 9*). Ordered states have been achieved for fibroblast cells near edges of microchannels (*2-4, 10*), on microgrooved substrates (*9, 11-14*), and on surfaces with stiffness gradients (*15, 16*). Of especial interests are substrates formed by liquid crystalline elastomers (LCEs), since these materials are themselves orientationally ordered and could dynamically change their properties in response to weak environmental cues (*17-23*).

In this work we present an approach to design orientationally ordered tissues with a predetermined spatially varying director $\hat{\mathbf{n}}(\mathbf{r})$ templated by the director pattern $\hat{\mathbf{n}}_{\text{LCE}}(\mathbf{r})$ of LCE substrates. Prior to polymerization, molecular orientation of the LCE precursor is patterned by the plasmonic photoalignment (*24*). The nanoscale molecular order of the LCE transcends to much longer length-scale to produce patterned living tissues formed by human dermal fibroblast cells (HDF). The tissue grows from the aqueous dispersion of the cells deposited onto the LCE substrate. Swelling in water produces anisotropic non-flat profile of the LCE that guides the growth of tissue. We demonstrate that the structured LCE imposes a dramatic effect on the tissues, by controlling the spatial distribution of cells, their density fluctuations and even their phenotype as evidenced by different cell aspect ratio. The patterned LCE pins the locations of topological defects in tissues through anisotropic surface interactions and limits unbinding of defect pairs. The dynamics of defects in cellular monolayer at the patterned substrates is similar to the dynamics of active nematics formed by extensile units and allows us to estimate elastic and surface anchoring parameters of the tissue. The study opens the possibility to engineer platforms for the controlled cell orientation and design it for the specific tissue (*25, 26*).

**Results and discussion**

The LCE is supported by a glass plate. To reduce surface roughness, the glass is covered with an indium tin oxide (ITO). The next layer is a photosensitive azo-dye, the in-plane alignment of which is patterned by light irradiation with spatially-varying linear polarization (*24*). This patterned azo-dye layer serves as a template for the monomer diacrylate RM257 doped with 5 wt% of photoinitiator I651 (Fig. S1). After alignment through the contact with the azo-dye, the monomer is UV-irradiated to photopolymerize the LCE substrate with a predesigned $\hat{\mathbf{n}}_{\text{LCE}}(\mathbf{r})$. Once the substrate is covered with an aqueous cell culture medium, the LCE swells. Within ~1 min, the LCE develops a non-flat profile with elongated ellipsoidal grains of an average height ~40 nm, length up to 30 μm and an aspect ratio higher than 2 (Figs. 1A, 2B, C and Fig. S2). Elongation of grains along $\hat{\mathbf{n}}_{\text{LCE}}$ is caused by the anisotropy of the elastic properties of the LCE and persist even when the director $\hat{\mathbf{n}}_{\text{LCE}}$ varies in space, Figs. 2-4. The dimension of grains is in the range that is sensed by the cells and is similar to the dimensions of grooves used in conventional lithography approaches (*8, 9*) and serve as a guiding rail for HDFs (Movie S1).

The orientational order of the uniformly aligned LCE substrate, calculated for grains longer than $10\,\mu\text{m}$, is high, $S_{\text{LCE}} = 0.99 \pm 0.01$ (see SM for the details of order parameter calculation). The order parameter $S_{\text{HDF}}$ of HDFs seeded on such a substrate, grows from $0.80 \pm 0.05$ to $0.96 \pm 0.01$ within the time interval 24-168 hours after seeding, before the tissue becomes confluent (Fig. 1B, C, Fig. S3A). Confluence, i.e. complete coverage of the substrate by the cells, occurs at concentration $\sigma_c \approx 3 \times 10^8 \text{ m}^{-2}$, after 240 hours from seeding. A strong increase of $S_{\text{HDF}}$ happens at $\sigma = (0.1 - 0.7)\sigma_c$, when the cells do not contact with each other (Fig. 1C). It implies that the orientational order is caused by the direct interactions of each cell with the LCE substrate. In the time interval 24-168 hours, the cells migrate in the plane of the substrates with the speed $\sim 10^2$ μm/hr (Movie S1). Development of orientational order in the HDF tissues is not associated with the evolution of the grainy texture of the substrate, since the latter, after the formation period of ~1 min, does not change over at least 5 days of observations. The LCE substrates mediate the alignment of both the cell bodies and elongated nuclei (Fig. 1D, E). The nuclei show a high order parameter reaching $S_{\text{nuclei}} = 0.88 \pm 0.05$ after 240 hours of cell seeding. The control ITO-glass substrates do not align cells, as $S_{\text{HDF}}, S_{\text{nuclei}} < 0.1$ (Fig. 1E, F, Fig. S3B-D).

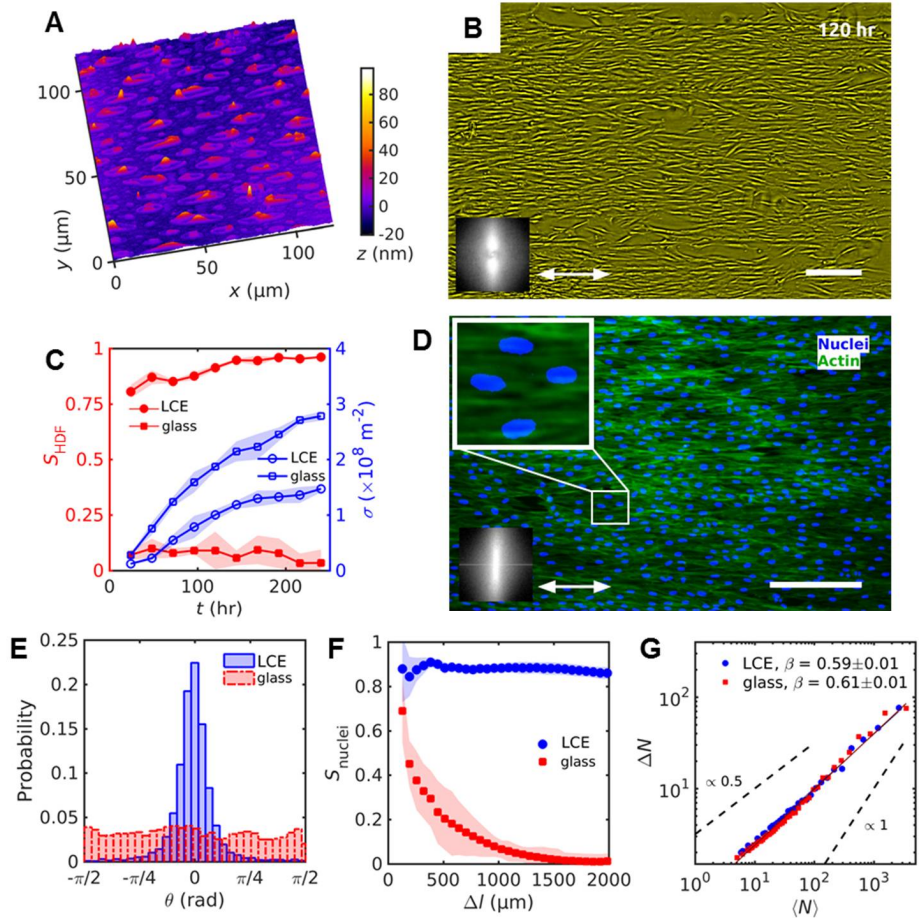

**Fig. 1.** Uniform alignment of HDF cells on LCE with a uniform $\hat{\mathbf{n}}_{LCE} = \text{const}$. (**A**) Digital holographic microscopy (DHM) texture of the LCE surface after contact with the aqueous growth medium. (**B**) Phase contrast microscopy (PCM) texture of HDF cells growing on LCE substrates at 120 hours after seeding. Double-headed arrow represents $\hat{\mathbf{n}}_{LCE}$. Insets show fast Fourier transformation of PCM textures indicating orientational order along the uniform $\hat{\mathbf{n}}_{LCE}$. (**C**) Evolution of the order parameter $S_{HDF}$ of cells bodies (filled red symbols) and cell density $\sigma$ (empty blue symbols). (**D**) Fluorescent microscopic textures of HDF cells on LCE; fluorescently labeled nuclei (blue) and cytoskeleton F-actin proteins (green). (**E**) Distribution of nuclei orientation. (**F**) Dependence of the order parameter $S_{nuclei}$ of nuclei on the size of a square subwindow. (**G**) Number density fluctuations $\Delta N$ calculated for the mean number of cell nuclei $\langle N \rangle$. All scale bars are 300 μm.

Both aligned and un-aligned assemblies of living cells exhibit high fluctuations of number density, Fig. 1G, which is an attribute of out-of-equilibrium systems (*27-29*). The dependency $\Delta N \sim \langle N \rangle^{\beta}$ of the standard deviation $\Delta N$ on the mean number of nuclei $\langle N \rangle$ calculated from fluorescent images in Fig. 1D following the procedure described in Refs. (*2, 29-31*), yields $\beta = 0.59 \pm 0.03$ for the aligned tissues and $\beta = 0.61 \pm 0.03$ for their un-aligned counterparts (Fig. 1G), higher than $\beta = 0.5$ expected in equilibrium. Both values are close to $\beta = 0.66 \pm 0.06$ measured for a misaligned array of mouse fibroblast cells (*2*).

The aligning ability of LCE substrates extends to spatially varying patterns with topological defects of charge $m = \pm 1/2$, $\pm 1$, ...., predesigned as

$$\hat{\mathbf{n}}_{\text{LCE}} = (n_x, n_y, n_z) = \left[ \cos(m\varphi + \varphi_0), \sin(m\varphi + \varphi_0), 0 \right], \quad (1)$$

where $\varphi = \arctan(y/x)$ and $\varphi_0 = \text{const}$; $m$ is a number of times the director reorients by $2\pi$ when one circumnavigates around the defect core. For pairs of defects of charge $m_1$ and $m_2$, separated by a distance $d_0$, the argument $m\varphi$ in Eq. (1) is replaced by $m_1 \arctan\left[ y/(x + d_0/2) \right] + m_2 \arctan\left[ y/(x - d_0/2) \right]$ (see SM for the explicit equations for different director patterns and Fig. S4). The spatial variation of the LCE director $\hat{\mathbf{n}}_{\text{LCE}}$ is imaged by the PolScope microscopy (see SM).

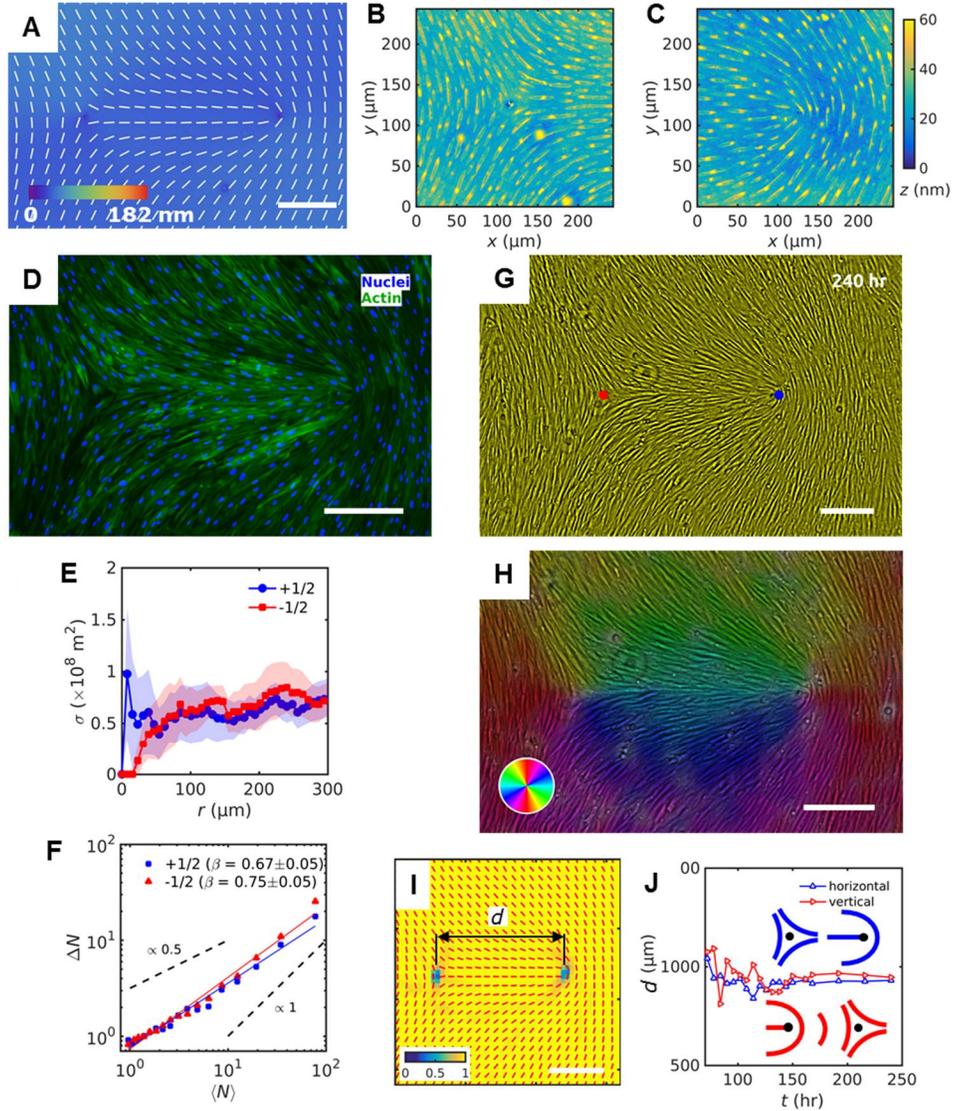

**Fig. 2.** Patterned alignment of HDF cells on LCE with a $(-1/2, +1/2)$ pair of defects. (**A**) PolScope texture showing $\hat{\mathbf{n}}_{\text{LCE}}$ and optical retardation of LCE in contact with the cell growth medium. (**B**, **C**) DHM textures of LCE surface in contact with the cell growth medium with $-1/2$ (**B**) and $+1/2$ (**C**) defects. (**D**) Fluorescently stained HDF cells; DAPI-labelled cells nuclei (blue) and phalloidin-labeled actin cytoskeleton filaments (green). (**E**) The surface density of cell nuclei $\sigma$ as the function of distance from $+1/2$ (blue) and $-1/2$ (red) defect cores. (**F**) Large number density fluctuations $\Delta N$ of the nuclei in the vicinity of defect cores. (**G**) PCM images of HDF cells on LCE substrate at 240 hours after the seeding. Blue and red dots denote location of $+1/2$ and $-1/2$ defect cores, respectively, obtained from polarizing optical microscopy (POM) texture of LCE. (**H**) Orientational field and (**I**) local nematic order parameter of HDF cells imaged with PCM. Red bars on (**I**) denote local orientation of cells long axis. (**J**) Separation between half-strength defects in horizontal and vertical (see Fig. S6) geometries. All scale bars are 300 μm.

The HDF cells self-organize into aligned assemblies that follow the pre-imposed director $\hat{\mathbf{n}}_{LCE}(\mathbf{r})$ (Figs. 2-5). The orientational order of the assemblies is apolar, $\hat{\mathbf{n}} = -\hat{\mathbf{n}}$, as evidenced by the presence of $+1/2$ and $-1/2$ disclinations, Fig. 2. The cores of these disclinations are co-localized with the cores of disclinations in the LCE (Fig. 2D, G, H, Fig. S5A, B, Fig. S6 and Movie S4). The director field of cells becomes poorly defined as one approaches the defect cores (Fig. 2H) with a corresponding decrease of $S_{HDF}$ (Fig. 2I), similar to the observation in (*3, 7*). In "wet" active nematics with momentum conservation, $+1/2$ and $-1/2$ defect pairs tend to unbind because of the high mobility of $+1/2$ defects (*32, 33*). Surface anchoring at the patterned LCE substrate prevents unbinding (Fig. 2J), thus overcoming the active and elastic forces (*28*). This surface anchoring prevails for relatively large separations between the defects ($d_0 = 1 \text{mm}$ in Figs. 2-4), but becomes comparable to elastic and active forces for shorter distances, as discussed later for the integer-strength defects.

The cells' concentration near the $+1/2$ cores is substantially higher than near $-1/2$ ones (Fig. 2E). The $\pm 1/2$ defects differ also in the characteristics of number density fluctuations: the fluctuations are stronger in the vicinity of $-1/2$ defects as compared to $+1/2$ defects (Fig. 2F), in both cases deviating from the exponents in equilibrium systems (*29, 34*).

The HDF cells alignment follows the patterned director of LCE that contains integer $+1$ defects of either pure splay (Fig. 3A, Fig. S5C and Movie S5), or bend (Fig. 4A, Fig. S5D and Movie S6). Azimuthal dependence of orientation $\theta$ of the HDF nuclei around $-1$ and $+1$ defects matches the predesigned $\hat{\mathbf{n}}_{LCE}(\mathbf{r})$ (Fig. 3B, E, F, Fig. 4B, E, F and Fig. S7) but only at some distance $r > (200 - 500) \, \mu\text{m}$ from the pre-inscribed defect core in $\hat{\mathbf{n}}_{LCE}$. The local cell density increases near the $+1$ cores (Fig. 3C). For example, at a distance $20 \, \mu\text{m}$ from the core, the $+1$ defects of a radial type show $\sigma_{r=20\,\mu\text{m}}^{(+1,\text{splay})} \approx 0.5 \times 10^8 \text{ m}^{-2}$, which is 1.5 times higher than the density far away from the core, at $r = 300 \, \mu\text{m}$, $\sigma_{r=300\,\mu\text{m}}^{(+1,\text{splay})} = 0.35 \times 10^8 \text{ m}^{-2}$ (Fig. 3C). Circular $+1$ defects exhibit even higher ability to concentrate cells: in Fig. 4C, $\sigma_{r=50\,\mu\text{m}}^{(+1,\text{bend})} \approx 1.5 \times 10^8 \text{ m}^{-2}$. In contrast, $-1$ defects deplete the density of the HDF cells: in Fig. 4C, $\sigma_{r=50\,\mu\text{m}}^{(-1,\text{bend})} \approx 0.5 \times 10^8 \text{ m}^{-2}$, which is 3 times lower than $\sigma_{r=50\,\mu m}^{(+1,\text{bend})}$ and 25-30% lower than the density far from the defect cores (Fig. 4C). Since the cells are in contact with each other, the strong variation of density causes dramatic changes in cell shapes. The aspect ratio of the HDF bodies near $+1$ circular defect is $2.6\pm1.5$ while near $-1$ defect it is much higher, $5.8\pm2.7$ (Fig. 4E and Fig. S8). Therefore, the LCE patterns influence the phenotype of the HDF cells in tissues. The fluctuations of number densities of cell nuclei also depend on the topological charge of the defects: $-1$ defects cause stronger fluctuations than $+1$ defects of both radial (Fig. 3D) and circular (Fig. 4D) geometries.

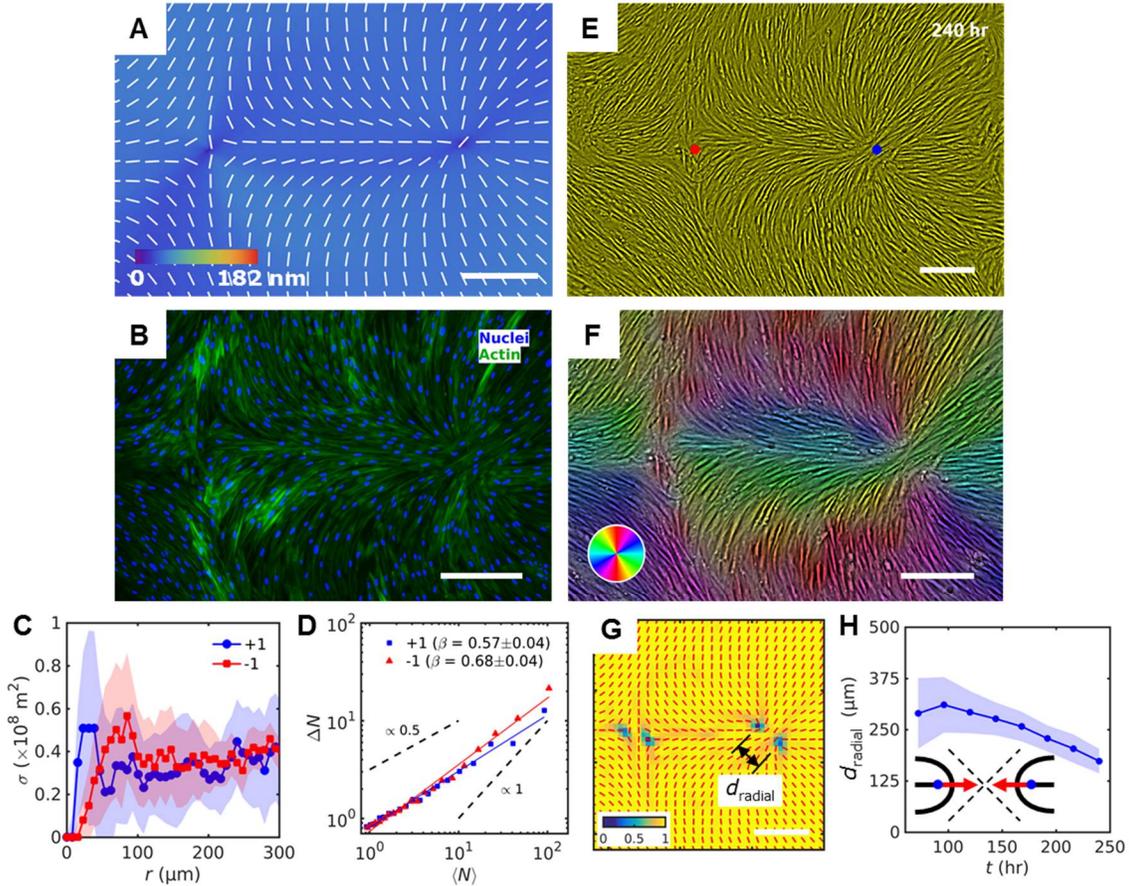

**Fig. 3.** Patterned alignment of HDF cells on LCE patterned with $(-1,+1)$ radial defects (splay type). (**A**) PolScope image of $\hat{\mathbf{n}}_{\text{LCE}}$ pattern of LCE in contact with the cell growth medium. (**B**) Fluorescently labeled HDF cells. (**C**) The surface density of cell nuclei $\sigma$ as the function of distance from defect cores. (**D**) Large number density fluctuations $\Delta N$ with mean number of nuclei $\langle N \rangle$ showing larger slope near $-1$ cores compare to $+1$. (**E**) PCM textures of HDF cells on LCE layer at 240 hours after cell seeding. Red and blue dots denote location of negative and positive sign defects in patterned LCE obtained from crossed polarized textures. (**F**) Orientational field and (**G**) local nematic order of HDF tissue obtained from local anisotropy of PCM texture in (**E**). Red bars on (**G**) denote local orientation of cells long axis. (**H**) Separation of $+1/2$ defects near the $+1$ radial core. All scale bars are 300 µm.

In an equilibrium 2D nematic, the defects of integer strength tend to split into pairs of semi-integer defects, since their elastic energy scale as $\propto m^2$ (*35*). Both $-1$ and $+1$ defects in the HDF tissues split into two defects of an equal semi-integer strength separated by a distance $d \approx (200-500)$ µm. Inside the region $r<d$, $\hat{\mathbf{n}}_{\text{HDF}}$ deviates significantly from $\hat{\mathbf{n}}_{\text{LCE}}(\mathbf{r})$. For the radial $+1$ configuration, the distance $d$ decreases with time, as the tissue grows towards confluency, while in the circular $+1$ case, $d$ increases with time (Fig. 4F-H). These very different separation scenarios are reproducible in multiple (more than ten) samples and are reminiscent of the defect dynamics in wet active nematics with extensile units, in which the comet-like $+1/2$ defects move with the "head" leading the way (*6, 7, 28, 36*). The splitting and motion of the half-integer defects is caused by the dynamics of the cells in the tissue rather than by changes of the underlying LCE geometry: the defects in LCE show no signs of splitting beyond about 1 µm over the entire duration of the experiment (see SM).

The data above demonstrate that the dynamics and proliferation of defects in patterned tissues can be arrested by the surface anchoring forces. In general, topological defects in sufficiently active matter tend to unbind and disorder the system (*32, 33*). The prepatterned tissue demonstrate that this defect unbinding can be prevented by interactions with an anisotropic patterned substrate which establishes a finite stationary $d$. Consider a pair of $+1/2$ disclinations at the substrate patterned as a $+1$ radial defect ($\hat{\mathbf{n}}_{\text{LCE}} = (n_r, n_\varphi, n_z) = (1,0,0)$ in cylindrical coordinates). In the so-called one-constant approximation, their elastic repulsive potential is weakly dependent on the separation distance $d$: $F_{\text{E}} = -\frac{\pi K h}{2} \ln \frac{d}{2r_c}$, where $K$ is the average Frank elastic modulus, $h \approx 20$ µm is the thickness of the cell layer, and $r_c$ is the size of the defect core (*35*). Activity, as proposed in (*32, 33*) contributes an additional force $f_a$ that tends to drive $+1/2$ defects either towards each other ($f_a < 0$) in the radial geometry or away from each other ($f_a > 0$) in the circular geometry. Since the defect splitting causes the tissue director $\hat{\mathbf{n}}_{\text{HDF}}$ to differ significantly from $\hat{\mathbf{n}}_{LCE}$ within an area of diameter $d$, the behavior of defects depends also on the surface energy penalty $F_S = \iint \frac{W}{2}\left[1-\left(\hat{\mathbf{n}}_{LCE} \cdot \hat{\mathbf{n}}_{HDF}\right)^2\right]dS = aWd^2$, where $a \approx 0.2$ is the numerical coefficient (see details in SM), $W$ is the azimuthal surface anchoring coefficient. The stationary value of the separation distance is then defined by minimizing

$$F = -\frac{\pi Kh}{2}\ln\frac{d}{2r_c} - f_a d + aWd^2, \qquad (2)$$

which yields $d_s = \left(f_a + \sqrt{f_a^2 + 4\pi ahKW}\right)/(4aW)$.

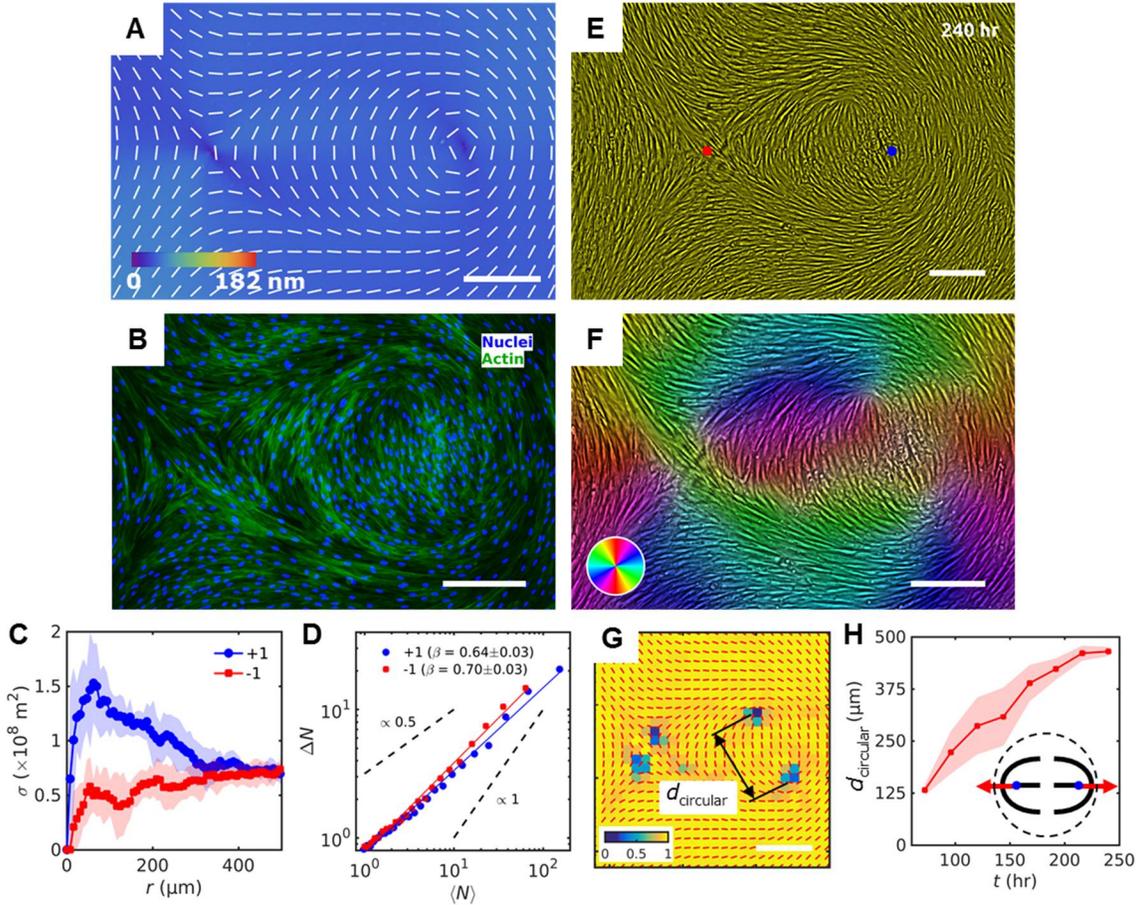

**Fig. 4.** Patterned alignment of HDF cells on LCE with pair of circular $(-1,+1)$ configuration (bend type). (**A**) PolScope image of $\hat{\mathbf{n}}_{LCE}$ pattern of LCE in contact with the cells growth medium. (**B**) Fluorescently stained HDF cells aligned in $(-1,+1)$ circular pattern. (**C**) Radial dependence of the surface density of cell nuclei $\sigma$ shows increase of cell density at $+1$ defect core. (**D**) Large number density fluctuations $\Delta N$ in cell nuclei calculated for increasing window size of regions with mean number of nuclei $\langle N \rangle$. $-1$ defects show larger slope compare to $+1$. (**E**) PCM image of HDF cells

on LCE layer at 240 hours after cell seeding. Red and blue dots denote location of negative and positive sign defect cores in patterned LCE obtained from POM. (**F**) Orientational field and (**G**) local nematic order of HDF tissue obtained from local anisotropy of the textures in (**E**). Red bars on (**G**) denote local orientation of cells long axis. (**H**) Separation distance between $+1/2$ defects near the $+1$ circular core. All scale bars are 300 µm.

According to Eq. (2), nonzero surface anchoring establishes a stationary finite separation of defects that is smaller in the radial geometry, and larger in the circular geometry, in agreement with the experiment, Figs 3G, H and 4G, H. The difference in stationary separations in circular and radial geometry caused by the opposite directionality of the active force $f_a$, $\Delta d_s = f_a/(2aW)$, allows us to estimate $f_a/W \approx 140\,\mu m$, since in the experiment, $\Delta d_s \approx 350\,\mu m$. We also observe splitting of $-1$ defects. The resulting $-1/2$ defects can be considered in first approximation as inactive (*32, 33*), thus the separation distance is expected to be $d_s = \sqrt{\pi hK/(4aW)}$; in the experiment, $d_s \approx 120\,\mu m$, thus the anchoring extrapolation length of the tissue is $b = K/W \approx 180\,\mu m$. Comparing this to the estimate $f_a/W \approx 140\,\mu m$, one concludes that the active force is on the order of the Frank elastic modulus of the tissue, $f_a \sim K$. Note that $b$ is comparable to the typical length of the fibroblast cell, $l \approx 100\,\mu m$, and suggests strong orientational anchoring by the LCE. The dependence of $W$ on systems parameters can be estimated by considering the non-flat profile of the substrates with a typical distance between the grains being $\lambda = 10\,\mu m$ and their height $u_0 = 0.1\,\mu m$, Fig. S2. When a cell is aligned orthogonally to the elongated grains, it must bend around the grain. Assuming the substrate profile to be sinusoidal, $u = u_0 \cos(2\pi x/\lambda)$, the bending energy density writes $\frac{K}{2}(\partial^2 u/\partial x^2)^2$ and the surface energy per unit area as $W \sim 4\pi^4 K h u_0^2/\lambda^4$ (see SM for details). $K$ can be estimated using the bending modulus of fibroblast cells, $\kappa = \frac{3E\pi h^4}{64l^3} \sim 2\times 10^{-4}$ N/m, where $E = 10^4$ Pa (*37*) is the typical Young's modulus, as $K \sim \kappa h \sim 5$ nN. The experimental value $b \approx 180\,\mu m$ then leads to $W = K/b \sim 2.5\times 10^{-5}$ J/m$^2$, which is close to the theoretical expectation $W \sim 4\pi^4 K h u_0^2/\lambda^4 \sim 3.7\times 10^{-5}$ J/m$^2$.

**Conclusion**

We demonstrate that LCE substrates with photopatterned structure of spatially varying molecular orientation can be used to grow biological tissues with predesigned alignment of cells. For the first time, human tissues are designed with predetermined type and location of topological defects in their orientationally ordered structure. Besides the alignment, the substrates affect the surface density of cells, fluctuations of the number density, and the phenotype of cells as evidenced by different aspect ratio of the cell bodies located in regions with different director deformations of the patterned tissue. In particular, higher density of cells is observed at the defect cores with positive topological charge, whereas the density is lower near the negative defects; cells are more round near positive defects and more elongated near negative defects. Anisotropic surface interactions between the tissue and the underlying LCE impose restrictions on the dynamics of topological defects in tissues, preventing uncontrolled unbinding. The mechanism of cell alignment is rooted in swelling of the substrates upon contact with the aqueous cell culture; swelling produces anisotropic non-flat topography that follows the predesigned photopatterned director field of the LCE. We estimated the anchoring strength of the LCE alignment layer and elastic modulus of the fibroblast tissues. The approach opens vast possibilities in designing biological tissues with predetermined alignment and properties of the cells, with precise location of orientational defects, which could lead to a controllable migration, differentiation, and apoptosis of cells. The proposed approach could be developed further by chemical functionalization of LCEs, by fabricating LCE substrates with dynamical topographies responsive to environmental cues, which would advance our understanding of the fundamental mechanisms underlying tissue development and regeneration.

**References and Notes:**
1. R. Kemkemer, V. Teichgräber, S. Schrank-Kaufmann, D. Kaufmann, H. Gruler, Nematic order-disorder state transition in a liquid crystal analogue formed by oriented and migrating amoeboid cells. *European Physical Journal E* **3**, 101-110 (2000).
2. G. Duclos, S. Garcia, H. G. Yevick, P. Silberzan, Perfect nematic order in confined monolayers of spindle-shaped cells. *Soft Matter* **10**, 2346-2353 (2014).
3. G. Duclos, C. Erlenkämper, J. F. Joanny, P. Silberzan, Topological defects in confined populations of spindle-shaped cells. *Nature Physics* **13**, 58-62 (2017).
4. G. Duclos *et al.*, Spontaneous shear flow in confined cellular nematics. *Nature Physics* **14**, 728-732 (2018).
5. H. Morales-Navarrete *et al.*, Liquid-crystal organization of liver tissue. *bioRxiv*, 1-18 (2018).

6. T. B. Saw *et al.*, Topological defects in epithelia govern cell death and extrusion. *Nature* **544**, 212-216 (2017).
7. K. Kawaguchi, R. Kageyama, M. Sano, Topological defects control collective dynamics in neural progenitor cell cultures. *Nature* **545**, 327-331 (2017).
8. D. H. Kim *et al.*, Mechanosensitivity of fibroblast cell shape and movement to anisotropic substratum topography gradients. *Biomaterials* **30**, 5433-5444 (2009).
9. J. M. Molitoris *et al.*, Precisely parameterized experimental and computational models of tissue organization. *Integrative Biology (United Kingdom)* **8**, 230-242 (2016).
10. N. D. Bade, R. D. Kamien, R. K. Assoian, K. J. Stebe, Edges impose planar alignment in nematic monolayers by directing cell elongation and enhancing migration. *Soft Matter* **14**, 6867-6874 (2018).
11. R. Kemkemer, S. Jungbauer, D. Kaufmann, H. Gruler, Cell orientation by a microgrooved substrate can be predicted by automatic control theory. *Biophysical Journal* **90**, 4701-4711 (2006).
12. D.-H. Kim *et al.*, Nanoscale cues regulate the structure and function of macroscopic cardiac tissue constructs. *Proceedings of the National Academy of Sciences* **107**, 565-570 (2009).
13. C. Rianna *et al.*, Reversible Holographic Patterns on Azopolymers for Guiding Cell Adhesion and Orientation. *ACS Applied Materials and Interfaces* **7**, 16984-16991 (2015).
14. C. Rianna *et al.*, Spatio-Temporal Control of Dynamic Topographic Patterns on Azopolymers for Cell Culture Applications. *Advanced Functional Materials* **26**, 7572-7580 (2016).
15. H. B. W. C. M. Lo, M. Dembo, and Y. L. Wang, Cell movement is guided by the rigidity of the substrate, vol. 79, no. 1, pp. 144–152, 2000. *Biophysical Journal* **79**, 144-152 (2000).
16. U. S. Schwarz, S. A. Safran, Physics of adherent cells. *Reviews of Modern Physics* **85**, 1327-1381 (2013).
17. A. M. Lowe, N. L. Abbott, Liquid crystalline materials for biological applications. *Chemistry of Materials* **24**, 746-758 (2012).
18. A. Agrawal *et al.*, Stimuli-responsive liquid crystal elastomers for dynamic cell culture. *Journal of Materials Research* **30**, 453-462 (2015).
19. G. Koçer *et al.*, Light-Responsive Hierarchically Structured Liquid Crystal Polymer Networks for Harnessing Cell Adhesion and Migration. *Advanced Materials* **29**, 1-8 (2017).
20. D. Martella *et al.*, Liquid Crystalline Networks toward Regenerative Medicine and Tissue Repair. *Small* **13**, 1-8 (2017).
21. A. Sharma *et al.*, Effects of Structural Variations on the Cellular Response and Mechanical Properties of Biocompatible, Biodegradable, and Porous Smectic Liquid Crystal Elastomers. *Macromolecular Bioscience* **17**, 1-14 (2017).
22. M. E. Prévôt *et al.*, Liquid crystal elastomer foams with elastic properties specifically engineered as biodegradable brain tissue scaffolds. *Soft Matter* **14**, 354-360 (2018).
23. D. Martella *et al.*, Liquid Crystal-Induced Myoblast Alignment. *Adv Healthc Mater* **8**, e1801489 (2019).
24. Y. Guo *et al.*, High-Resolution and High-Throughput Plasmonic Photopatterning of Complex Molecular Orientations in Liquid Crystals. *Advanced Materials* **28**, 2353-2358 (2016).

25. A. Lee *et al.*, 3D bioprinting of collagen to rebuild components of the human heart. *Science* **365**, 482-487 (2019).
26. B. Grigoryan *et al.*, Multivascular networks and functional intravascular topologies within biocompatible hydrogels. *Science* **364**, 458-464 (2019).
27. S. Ramaswamy, R. A. Simha, J. Toner, Active nematics on a substrate: Giant number fluctuations and long-time tails. *Europhysics Letters* **62**, 196-202 (2003).
28. M. C. Marchetti *et al.*, Hydrodynamics of soft active matter. *Reviews of Modern Physics* **85**, 1143-1189 (2013).
29. V. Narayan, S. Ramaswamy, N. Menon, Long-lived giant number fluctuations in a swarming granular nematic. *Science* **317**, 105-108 (2007).
30. H. Chaté, F. Ginelli, R. Montagne, Simple model for active nematics: Quasi-long-range order and giant fluctuations. *Physical Review Letters* **96**, 1-4 (2006).
31. D. Nishiguchi, K. H. Nagai, H. Chaté, M. Sano, Long-range nematic order and anomalous fluctuations in suspensions of swimming filamentous bacteria. *Physical Review E* **95**, 020601 (2017).
32. L. Giomi, M. J. Bowick, X. Ma, M. C. Marchetti, Defect annihilation and proliferation in active Nematics. *Physical Review Letters* **110**, 1-5 (2013).
33. S. Shankar, S. Ramaswamy, M. C. Marchetti, M. J. Bowick, Defect Unbinding in Active Nematics. *Physical Review Letters* **121**, 108002 (2018).
34. R. Aditi Simha, S. Ramaswamy, Hydrodynamic fluctuations and instabilities in ordered suspensions of self-propelled particles. *Phys Rev Letters* **89**, 058101 (2002).
35. M. Kleman, O. D. Lavrentovich, *Soft Matter Physics: An Introduction*. (Springer-Verlag New York, Inc., New York, N. Y., 2003).
36. T. B. Saw, W. Xi, B. Ladoux, C. T. Lim, Biological Tissues as Active Nematic Liquid Crystals. *Advanced Materials* **1802579**, 1-12 (2018).
37. T. G. Kuznetsova, M. N. Starodubtseva, N. I. Yegorenkov, S. A. Chizhik, R. I. Zhdanov, Atomic force microscopy probing of cell elasticity. *Micron* **38**, 824-833 (2007).

**Acknowledgments:** We thank Prof. Irakli Chaganava for synthesizing the photoalignment material SD1. **Funding:** The work is supported by NSF DMREF grant DMS-1729509 and by Office of Sciences, DOE, grant DE-SC0019105. **Author contributions:** T.T. and O.D.L. conceived the research, O.D.L. supervised the project, T.T. designed and performed the experiments and imaging, J.K. prepared cell culture, G.B. assisted in the immunofluorescent preparation, H.Y. and Q.-H.W. provided the plasmonic photomask for patterned photoalignment, T.T., S.V.S., and O.D.L. developed computational model, M.-H.K. coordinated the cell culture preparation, T.T. and O.D.L. analyzed the data and wrote the manuscript (with the input from all coauthors). **Competing interests:** The authors declare no competing interests. **Data and materials availability:** All data (including supporting movies) in support of the reported findings and computer code are available from the corresponding authors upon reasonable request.

Supplementary Materials for

**Topology control of human cells monolayer by liquid crystal elastomer**

Taras Turiv*, Jess Krieger, Greta Babakhanova, Hao Yu, Sergij V. Shiyanovskii, Qi-Huo Wei, Min-Ho Kim, Oleg D. Lavrentovich*

*Correspondence to: tturiv@kent.edu, olavrent@kent.edu

**Materials and methods**

*Substrate preparation and patterned alignment*
We use indium tin oxide (ITO) coated glass slides of the rectangular shape and 15 mm × 12 mm size. The glass is cleaned in the ultrasonic bath of deionized water and small concentration of detergent at 65°C for 20 minutes. Next, we rinse the glass with deionized water and isopropanol and let it dry for 10 min, after this the glass is further cleaned inside the UV-ozone chamber for 15 minutes to remove any remaining organic contamination and improve the wettability properties. Immediately after that the glass is spincoated with 0.5 wt% azo-dye SD-1 (synthesized following (*37*)) solution in dimethyl formamide, which serves as the photoalinging layer and annealed inside the oven at 100°C for 30 minutes. The thickness of the azo-dye layer is about 10 nm (*38*). For some samples we use azo-dye Brilliant Yellow (Sigma), no difference in the alignment of the HDF cells and quality of the film is observed. We illuminate the glass with metal-halide X-Cite 120 lamp through linear polarizer to create uniform alignment (easy axis of alignment is perpendicular to the linear polarization of light) or through the special photomask (*39*), which gives the possibility to create high resolution patterns of nonuniform director field. The illumination is performed for 5 minutes. The glass with azo-dye coating is spincoated with 6.65 wt% LC diacrylate monomer RM257 (Wilshire Technologies) and 0.35 wt% of photoinitiator I651 (Ciba Specialty Chemicals Inc.) in 93 wt% of toluene and polymerize it with 365 nm UV light of UVP-58 hand held lamp for 15 minutes (light intensity $1.8\,\mathrm{mW\,cm^{-2}}$) at room temperature and ambient atmosphere. After full solvent evaporation the concentration of photoinitiator I651 becomes 5 wt%. We calculated the thickness of LCE layer from the average optical retardance $\Gamma = 40$ nm measured with PolScope (fon the uniform director field). Since the birefringence of RM257 is about $\Delta n \approx 0.2$ (*40*) the thickness of LCE layer is $\Gamma/\Delta n \approx 200$ nm. The substrate is placed into a Petri dish, additional sterilization and cleaning is done inside UV-ozone chamber and then the cells are seeded onto the substrate and allowed to adhere, grow and multiply on it. LCE substrates with fibronectin extracellular matrix coating, used to improve the adhesion of HDF cells, are prepared by immersing LCE substrates into the solution of 1 wt% fibronectin (Thermo Fischer) in deionized water and incubated for 2 hours. Similar to fibrobnectin-free LCE substrates HDF cell alignment with uniform director of LCE with somewhat smaller order parameter $S_{\mathrm{HDF}}$, Fig. S4.

*Cell plating*
HDF cells were purchased from ATCC (Cat # PCS-201-010) and maintained at 37°C with 5% $CO_2$ and >90% humidity inside the incubator. Passage 1 to 6 are used in the experiment and no significant difference in the behavior of different passages is recorded. Cell culture medium is composed of Dulbecco's modified Eagle's medium (DMEM, high glucose) supplemented with 10% fetal bovine serum (FBS) (Clonetics), Glutamax, and penicillin/streptomycin (both from Thermo Fischer). The cells were plated at the surface density $3.3\times10^7$ m$^{-2}$ (fixed for all experiments) onto the substrates placed in a Petri dish made of a sterilized glass (to allow for the POM imaging).

*Immunocytochemistry*
F-actin cytoskeleton and nuclei in fibroblast cell cultures are immunofluorostained. Cell cultures are fixed in 4% PFA for 20 min, permeabilized with 0.1% Triton-X in phosphate buffer saline (PBS) for 25 min, and nonspecific enzymatic activity was blocked with 5% FBS in PBS for 20 min. Cultures are then incubated with Alexa Fluor 488 Phalloidin (1:100 dilution, A12379, Invitrogen) for 30 min and DAPI (D1306, Invitrogen) for 10 min.

*Microscopic observation*
The cells are examined under inverted phase contrast microscopes AmScope AE2000 or Leica DH (equipped with AmScope MU-2003-BI-CK or Pixelink PL-P755CU-T cameras, pair of polarizers, 4×/Ph0 or 10×/Ph1 objectives) without temperature and environmental control for short period of time (not more than 5 minutes) every 6 or 24 hours and kept inside the incubated rest of the time until cells reached confluency. At that point cells nuclei and cytoskeleton actin filaments are stained with the fluorescent dyes (see Methods) and observed under fluorescent microscope Nikon TE-2000i with sets of fluorescent cubes that allow excitation of different fluorescent dyes. The fluorescent data were recorded with Emergent HS-20000C camera with 10× and 20× long working distance objectives. The surface characterization was performed using atomic force microscopy (AFM) (AFMWorkshop), scanning electron microscopy (Quanta 450 FEG SEM/EBL) and digital holographic microscopy in reflection mode (Lyncee DHM R-1000). Imaging with the fluorescent microscope Nikon TE-2000i allows us to visualize elongated shapes of the cell nuclei and to determine $S_{\text{nuclei}}$ as described below. PolScope microscopy is used for birefringent materials, such as orientationally ordered LCE, and allows one to determine the local orientation of the in-plane optic axis (which coincides with the director $\hat{\mathbf{n}}_{\text{LCE}}$) and optical retardance (*41*). The technique is based on retardance mapping of the birefringence sample by passing a circularly polarized light through the sample and tunable liquid crystal compensator. The retardance map is reconstructed from the resulting image obtained by the microscope camera.

*Image analysis*
Images are processed in opensource software package Fiji/ImageJ2 (*42*), further analysis was performed with the custom written code in MATLAB (MathWorks) and Mathematica (Wolfram) proprietary packages. The local cell director $\hat{\mathbf{n}}_{\text{HDF}}$ is obtained from the phase contrast microscopic images and computed using OrientationJ ImageJ/Fiji plugin (*43*), which calculates local orientational tensor using finite difference intensity gradient method within the small interrogation window. The minimum window size was chosen to be of the order of typical isolated HDF cell width on the LCE substrates which is 20 µm. The long axis orientation and position of the cells' DAPI-stained nuclei $\hat{\mathbf{n}}_{\text{nuclei}}$ is obtained from fluorescent microscopic images. The orientation and intensity-weighted center of mass of each individual nucleus is calculated with Extended Particle Analyzer BioVoxxel ImageJ/Fiji plugin (*44*). Analysis of the azimuthal distribution of orientation and radial distribution of HDF cell nuclei density around topological defects, density fluctuation is performed using custom written program in MATLAB package. Radial density $\sigma$ is calculated as the ratio between number of nuclei and the area of the radial annulus of width 10 µm at the given radius $r$ from the center of +1 (blue line) and −1 (red line) defect cores in LCE.

**Supplementary text**

*Director field design*
Equation (1) slows how to produce director patterns with different charges of topological defects. We present the concrete expressions and the director maps that correspond to ones used to prepare substrates in Figs. 2-4.

- Pair of −1/2 and +1/2 defects is described by Eq. (1) with $m\varphi = m_1 \arctan\left[y/(x+d_0/2)\right] + m_2 \arctan\left[y/(x-d_0/2)\right]$, where $m_1 = -0.5$, $m_2 = 0.5$, $d_0 = 1$ mm and $\varphi_0 = \pi/2$, for the pair with the director parallel to the line connecting two defect cores, or $\varphi_0 = 0$, for the pair with the director parallel to the line connecting two defect cores (Fig. S5A, B).
- Pair of −1 and radial +1 defects is described by Eq. (1) with $m\varphi = m_1 \arctan\left[y/(x+d_0/2)\right] + m_2 \arctan\left[y/(x-d_0/2)\right]$, where $m_1 = -1$, $m_2 = 1$, $d_0 = 1$ mm and $\varphi_0 = 0$ (Fig. S5C).
- Pair of −1 and radial +1 defects is described by Eq. (1) with $m\varphi = m_1 \arctan\left[y/(x+d_0/2)\right] + m_2 \arctan\left[y/(x-d_0/2)\right]$, where $m_1 = -1$, $m_2 = 1$, $d_0 = 1$ mm and $\varphi_0 = \pi/2$ (Fig. S5D).

*Determination of scalar order parameter S*
The measure of two-dimensional long-range orientational order in a system of elongated LCE grains ($S_{\text{LCE}}$), HDF cells body ($S_{\text{HDF}}$) and elongated HDF nuclei ($S_{\text{nuclei}}$) is the tensor order parameter $Q_{ij} = 2\langle u_i u_j \rangle - \delta_{ij}$. Here the unit vector $\mathbf{u}$, with in-plane components $u_i$ and $u_j$ taken in the dyadic product, characterizes the axis along which an individual LCE grain, cell body or cell nucleus is elongated; $\delta_{ij}$ is the identity matrix, $\langle...\rangle$ means averaging over all cells (or nuclei) within the field of view of size $w \times h = 4\,\text{mm} \times 3\,\text{mm}$. The maximum eigenvalue of $Q_{ij}$ yields the scalar order parameter $S_{\text{LCE}}$ for LCE grains, $S_{\text{HDF}}$ for HDF cells or $S_{\text{nuclei}}$ for HDF cell nuclei. The distribution of local nematic order parameter was calculated from the cell orientational field $\hat{\mathbf{n}}_{\text{HDF}}$ by calculating maximum eigenvalue of nematic tensor order parameter within moving square window of the side size 60 µm.

*Dynamics of cells vs. fixed geometry of the LCE substrates*
The dynamics of cells that manifests itself in splitting of integer strength defects and in in-plane motion of +1/2 defects described in the main text, Figs. 3,4, is the intrinsic feature of the living tissues that is not related to the potential evolution of the LCE substrate. The latter is fixed as specified by Eq. (1) and remains intact over the entire duration of the experiment. Figure S9 illustrate the statement by showing side-by-side images of the LCE substrate and of the tissue grown at this substrate. The LCE is imaged by POM; under POM, the cells are not visible. The tissue is imaged by PCM, which is is not sensitive to light polarization. As clearly seen from co-localization of the textures, the dynamics of tissue defects is not associated with the dynamics of the LCE substrate.

*Surface anchoring energy calculation*

To obtain total surface free energy of the aligning array of fibroblast cells on LCE we integrated the surface anchoning energy per unit area $\frac{W}{2}\left[1-(\hat{\mathbf{n}}_{\text{LCE}} \cdot \hat{\mathbf{n}}_{\text{HDF}})^2\right]$ in the Rapini-Popular form in the polar coordinate system inside the region of the size $d$:

$$F_{\text{S}} = \int_0^{2\pi}\int_0^d \frac{W}{2}\left[1-(\hat{\mathbf{n}}_{\text{LCE}} \cdot \hat{\mathbf{n}}_{\text{HDF}})^2\right] r\, dr\, d\theta = \frac{Wd^2}{32}\int_0^{2\pi}\left[5-\sqrt{25-16\cos^2\theta}\right]d\theta = \frac{Wd^2}{16}\left(5\pi - 10E(16/25)\right),$$

where $E(m)$ is the complete elliptic integral of the second kind *(45),* which is evaluated to obtain $F_{\text{S}} = aWd^2$ with $a = \frac{1}{16}\left(5\pi - 10E(16/25)\right) \approx 0.2$.

**References**

37. H. Akiyama *et al.*, Synthesis and properties of azo dye aligning layers for liquid crystal cells. *Liquid Crystals* **29**, 1321-1327 (2002).
38. J. Wang *et al.*, Effects of humidity and surface on photoalignment of brilliant yellow. *Liquid Crystals* **44**, 863-872 (2017).
39. Y. Guo *et al.*, High-Resolution and High-Throughput Plasmonic Photopatterning of Complex Molecular Orientations in Liquid Crystals. *Advanced Materials* **28**, 2353-2358 (2016).
40. H. Ren, S. Xu, Y. Liu, S.-T. Wu, Switchable focus using a polymeric lenticular microlens array and a polarization rotator. *Optics Express* **21**, 7916 (2013).
41. M. Shribak, R. Oldenbourg, Techniques for fast and sensitive measurements of two-dimensional birefringence distributions. *Applied Optics* **42**, 3009 (2003).
42. J. Schindelin *et al.*, Fiji: An open-source platform for biological-image analysis. *Nature Methods* **9**, 676-682 (2012).
43. R. Rezakhaniha *et al.*, Experimental investigation of collagen waviness and orientation in the arterial adventitia using confocal laser scanning microscopy. *Biomechanics and Modeling in Mechanobiology* **11**, 461-473 (2012).
44. J. Brocher, Qualitative and Quantitative Evaluation of Two New Histogram Limiting Binarization Algorithms. *International Journal of Image Processing (IJIP)*, 30-48 (2014).
45. M. Abramovitz, I. A. Stegun, *Handbook of mathematical functions: with formulas, graphs, and mathematical tables*. (United States Department of Commerce, National Bureau of Standards; Dover Publications, Washington D. C.; New York, N. Y., 1972), vol. 17, pp. 590.

**Supplementary figures**

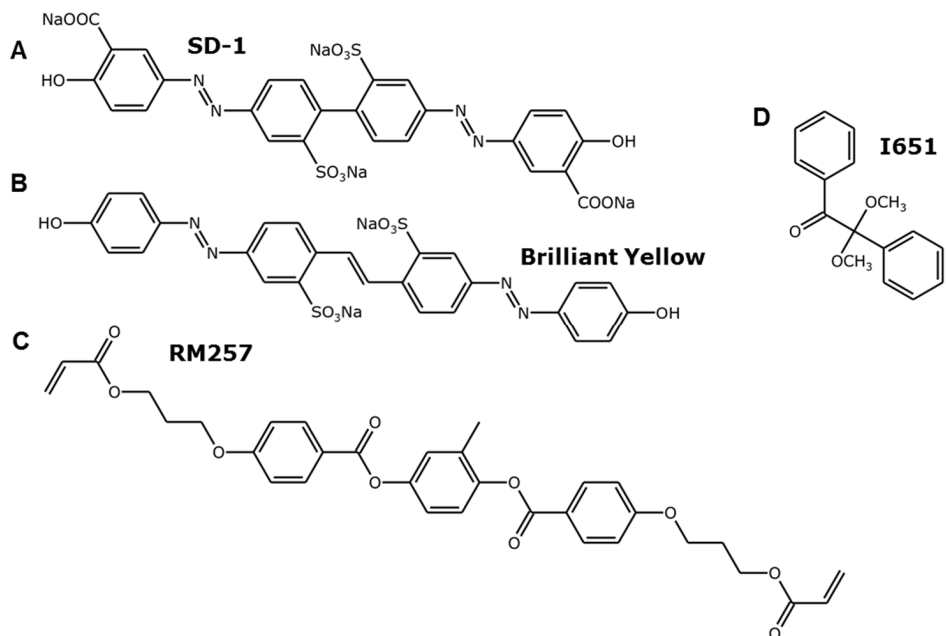

**Fig. S1.** Chemical structures of substances used. Chemical structure of (**A**) photoaligning azo-dye SD-1 and (**B**) Brylliant Yellow, (**C**) LCE diacrylate RM257 which is the main component of alignment layer for the tissues of HDF cells and (**D**) photoinitiator I651.

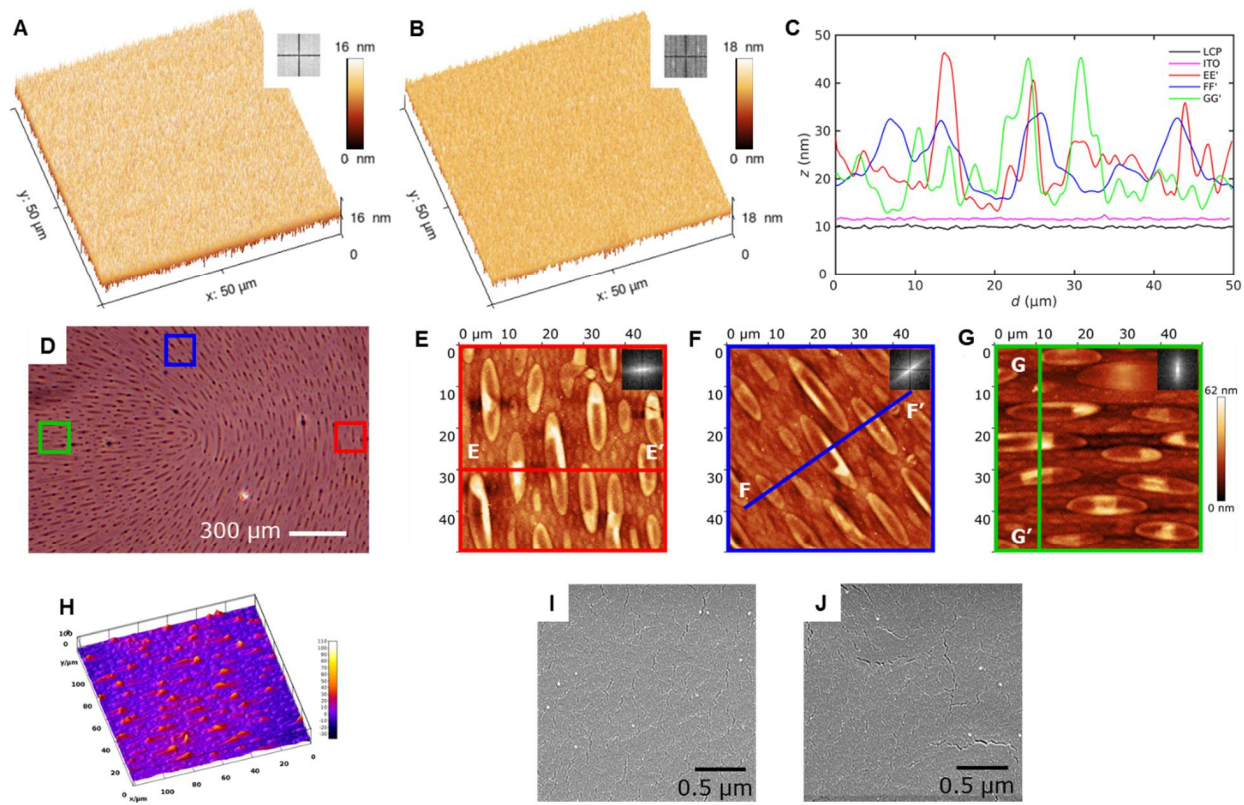

**Fig. S2.** AFM scans of surface used for proliferation of HDF cells in vitro. (**A**) LCE film with the uniform director along $x$-axis and (**B**) clean ITO glass before immersion of the glass into the cell culture medium. Insert textures represent 2D fast Fourier transformation of height map obtained from AFM with maximum spatial periodicity $0.5\,\mu m^{-1}$. (**C**) Profile of the surfaces before and after immersion. Surface profile taken along $x$-axis at $y = 12.5\,\mu m$ showing mean roughness of both LCE and ITO glass to be less than 2 nm. (**D**) Phase contrast microscopic image of patterned surface with topological defect $+1/2$ in LCE after the substrate was exposed to cell culture medium for 24 hours and dried for 3 hours at room temperature. (**E-G**) AFM scan of the substrate in (**D**) for different regions shown with red, blue and green squares. (**H**) Digital holographic microscopy texture of LCE with uniform director. Scanning electron microscopy of substrates with LCE on (**I**) photoalignment azo-dye layer and (**J**) clear ITO glass.

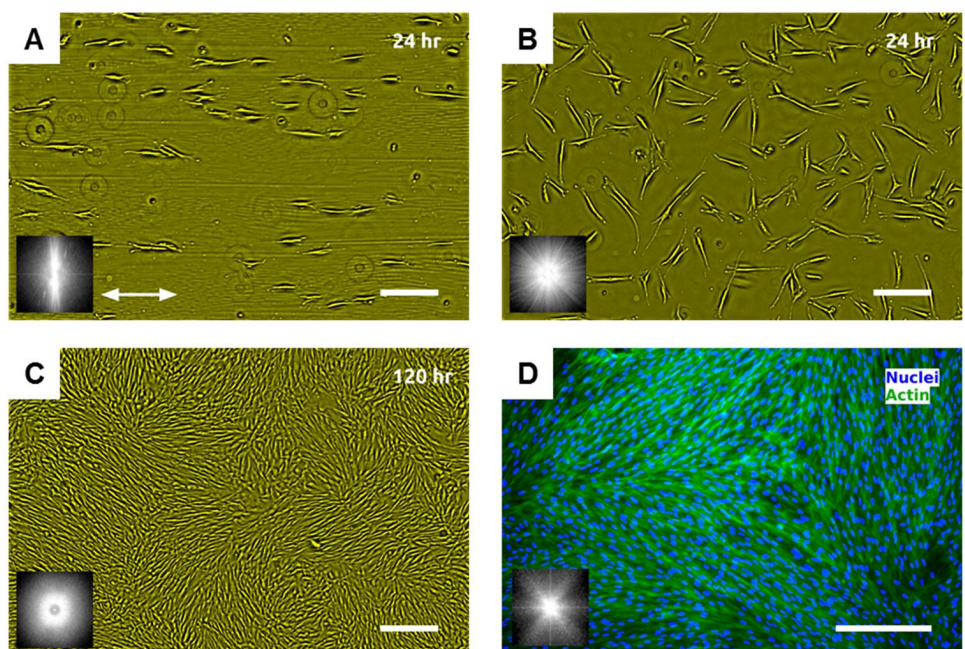

**Fig. S3.** HDF cells on LCE and clean glass substrates. (**A**-**C**) PCM textures of HDF cells grown on (**A**) LCE with the uniform director field along horizontal axis (white double arrow) at 24 hrs after seeding and on clean ITO glass at (**B**) 24 hours (**C**) and 120 hrs after seeding. (**D**) Fluorescent microscopic textures of HDF cells at the confluency (240 hrs) glass with fluorescently labeled nuclei (blue) and cytoskeleton F-actin proteins (green). All scale bars are 300 µm.

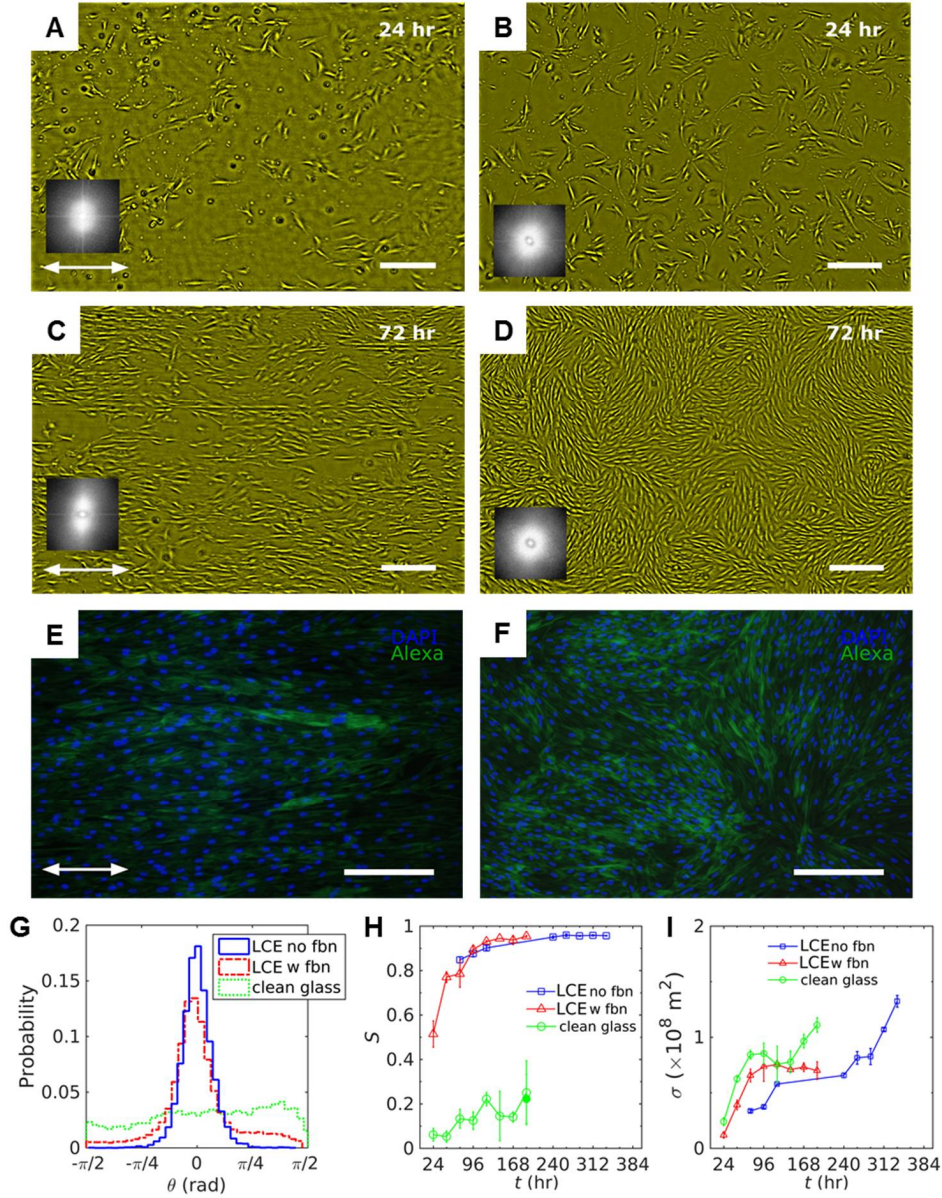

**Fig. S4.** Uniform alignment of migrating and proliferating HDF cells on uniaxially aligned LCE layer at different observation stages with fibronectin extracellular matrix coating. (**A**-**D**) PCM imaging of HDF cells growth at (**A**) 24 and (**C**) 72 hours after cells seeding on uniformly aligned LCE and at (**B**) 24 and (**D**) 72 hours after cells seeding on clean ITO glass. Both surfaces are precoated with fibronectin before cell seeding with the initial density of $3.3 \times 10^7$ m$^{-2}$. (**E**, **F**) Fluorescently stained HDF cells on (**E**) LCE and (**F**) ITO glass. (**G**) Distribution of the nuclei orientation. (**H**) Scalar order parameter $S_{HDF}$ calculated from the cell's long axis orientation obtained from PCM textures. (**I**) Surface density of the HDF cells. All scale bars are 300 μm.

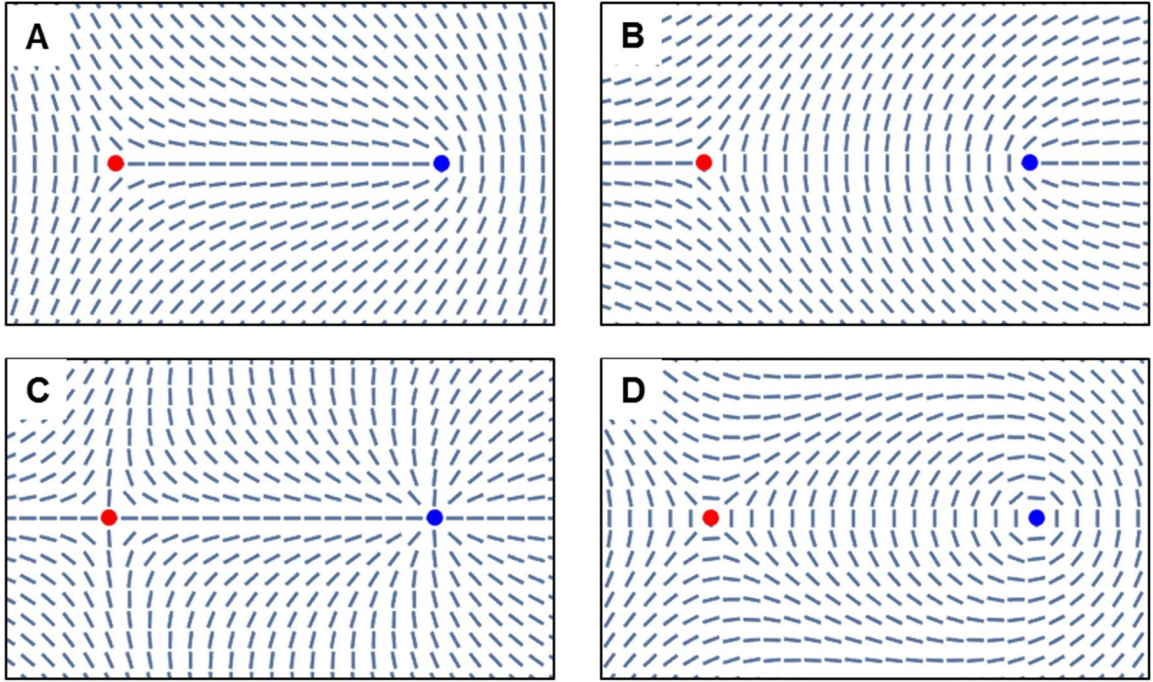

**Fig. S5.** The designed $\hat{\mathbf{n}}_{LCN}$ containing topological defects. (**A**) Pair of $+1/2$ (blue disk) and $-1/2$ (red disk) defects with the director parallel to the line connecting them. (**B**) Pair of $+1/2$ (blue disk) and $-1/2$ (red disk) defects with the director perpendicular to the line connecting them. (**C**) Pair of radial type $+1$ (blue disk) and $-1$ (red disk) defects. (B) Pair of circular type $+1$ (blue disk) and $-1$ (red disk) defects.

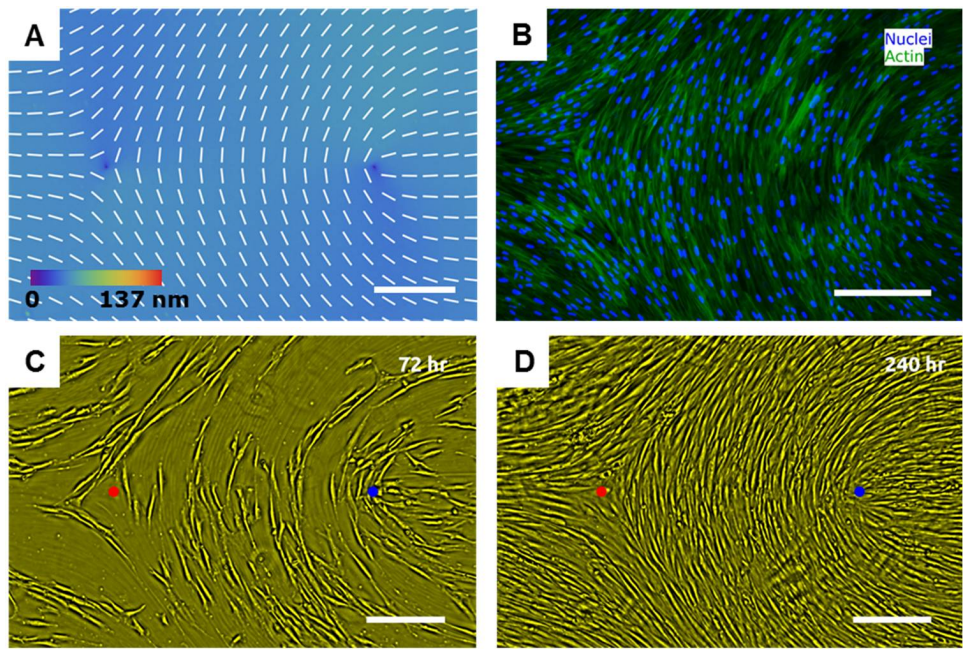

**Fig. S6**. HDF cells aligned in the pattern of half-strength defects with the director perpendicular to the line connecting two cores. (**A**) Director field of LCE obtained from PolScope. (**B**) Fluorescently stained nuclei (DAPI) and cytoskeleton F-actin filaments of the HDF cells. (**C, D**) PCM imaging of HDF cells proliferating on LCE at (**C**) 72 and (**D**) 240 hours of observation. Cells are seeded with the density of $3.3 \times 10^7$ m$^{-2}$. All scale bars are 300 μm.

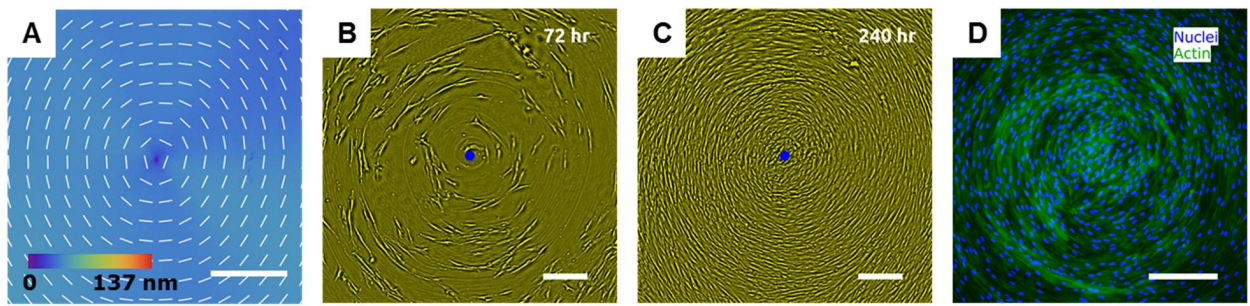

**Fig. S7.** HDF cells on patterned LCE layer with +1 circular defect core. Director field around +1 circular defect core of LCE obtained from PolScope (A). Phase contrast imaging of HDF cells growth at 72 and 240 hours after cells are seeded with the density of $3.3 \times 10^7$ m$^{-2}$ (B, C) and fluorescently labeled (D) nuclei (DAPI) and cytoskeleton (phalloidin). All scale bars are 300 μm.

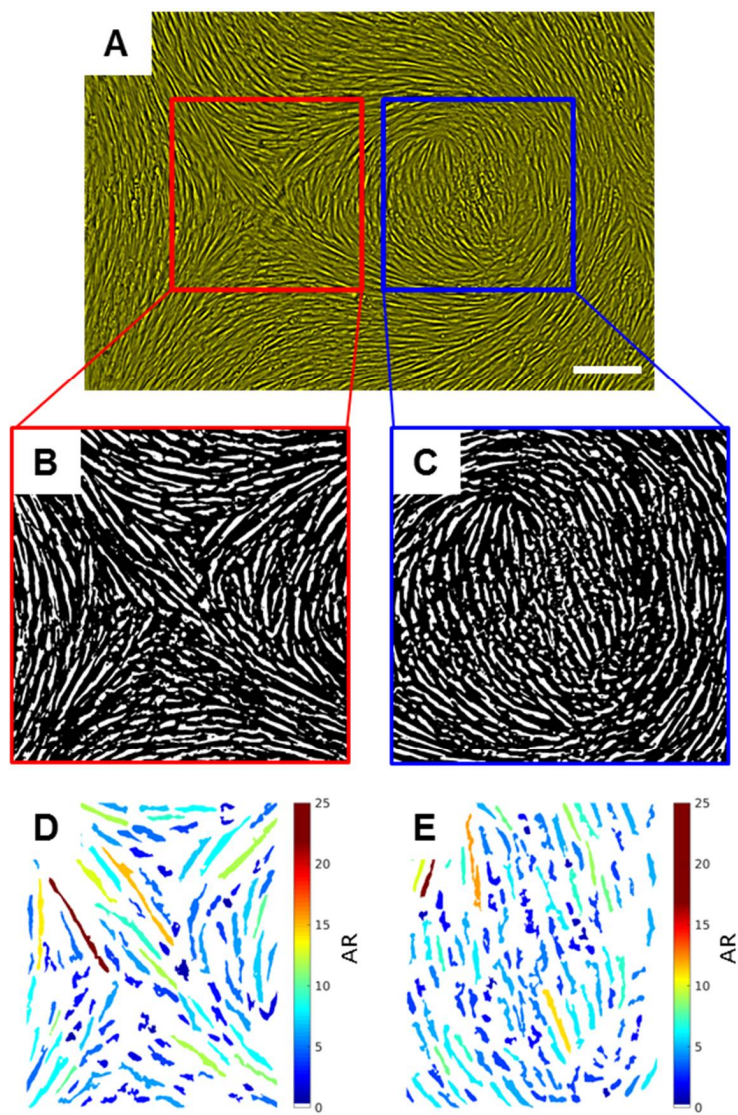

**Fig. S8**. HDF cells phenotype difference on LCE layer with $(+1,-1)$ circular defects pair. (**A**) PCM texture of HDF cells. (**B**, **C**) Images of intensity thresholded PCM textures of HDF cells grown near (**B**) −1 and (**C**) +1 defect. (**D**, **E**) Colorcoded maps of intensity thresholded images showing the cells of different aspect ratio (AR). The average AR over all cells near (**D**) −1 defect is $5.8 \pm 2.7$ and (**E**) +1 defect is $2.6 \pm 1.5$ (error is a standard deviation). The scale bar is 300 μm.

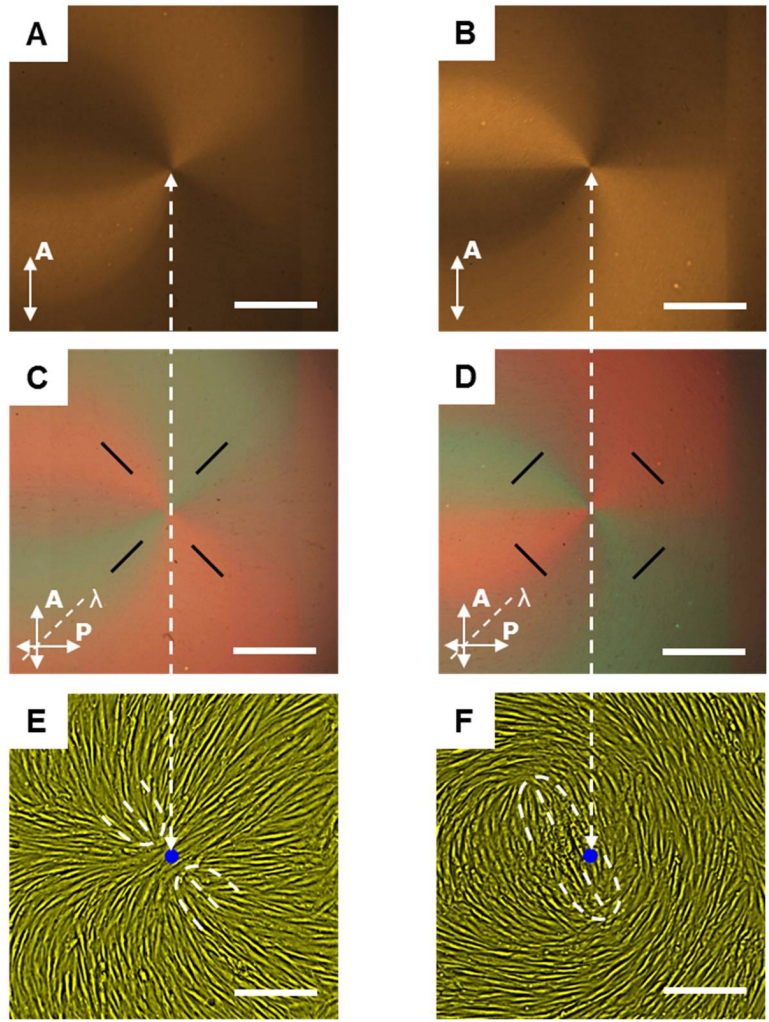

**Fig. S9**. Collocalization of topological defects in LCE with defects created in active layer of living HDF cells grown on top. (**A**) POM texture of +1 radial and (**B**) +1 circular type defects, image with analyzer inserted (polarization axis orientation is denoted as double-arrow line). The core of the defects is seen as the intersection of bright and dark brushes (white dashed double-arrow line). (**C**) POM texture of +1 radial and (**D**) +1 circular type defects, image with polarizer, analyzer, and full-wave retardation plate inserted (the slow optical axis of the retarder is denoted with dashed line). Black bars represent the orientation of the LCE director far from the defect core: blue-green color represents the optical retardation addition with the local $\hat{\mathbf{n}}_{LCE}$ being parallel to the slow axis of the retarder and red-orange color regions represent local $\hat{\mathbf{n}}_{LCE}$ that is perpendicular. PCM textures of HDF cells at 240 hours after seeding with the density of $3.3 \times 10^7$ m$^{-2}$ on LCE with (**E**) radial +1 and (**F**) circular +1 defect. Dashed white lines represent the local HDF cell orientation $\hat{\mathbf{n}}_{HDF}$ with two half-strength defects. Blue disks represent the defects cores of LCE colocalized from POM textures. All scale bars are 300 μm.

27